\documentclass[12pt]{article}
\usepackage[toc,page]{appendix}
\usepackage{graphicx}
\usepackage{float}
\usepackage{apacite}
\usepackage{hyperref}
\hypersetup{
	colorlinks,
	citecolor=black,
	filecolor=black,
	linkcolor=black,
	urlcolor=black
}
\usepackage{adjustbox}
\usepackage{titletoc}
\usepackage{booktabs}
\usepackage{ntheorem}
\usepackage{bbm}
\usepackage{amssymb,amsmath,amsfonts,eurosym,geometry,ulem,graphicx,caption,color,enumerate,moresize,setspace,sectsty,comment,footmisc,caption,natbib,pdflscape,subfigure,array,hyperref,natbib,bbm,comment,titling}
\usepackage[utf8]{inputenc}
\usepackage[english]{babel}
\usepackage{comment}
\usepackage{xparse}
\usepackage{amsmath}
\usepackage{hyperref}

\NewDocumentCommand{\multicite}{ogogogog}{%
	\citetext{%
		\IfValueT{#2}{%
			\IfValueT{#1}{\citealp[p.~#1]{#2}}%
			\IfNoValueT{#1}{\citealp{#2}}%
		}%
		\IfValueT{#4}{%
			;
			\IfValueT{#3}{\citealp[p.~#3]{#4}}%
			\IfNoValueT{#3}{\citealp{#4}}%
		}%
		\IfValueT{#6}{%
			;
			\IfValueT{#5}{\citealp[p.~#5]{#6}}%
			\IfNoValueT{#5}{\citealp{#6}}%
		}%
		\IfValueT{#8}{%
			;
			\IfValueT{#7}{\citealp[p.~#7]{#8}}%
			\IfNoValueT{#7}{\citealp{#8}}%
		}%
	}
}

\newtheorem{corollary}{Corollary}

\newtheorem{lemma}{Lemma}

\newtheorem{proposition}{Proposition}

\newenvironment{App-Proof}[1]{\textbf{Proof of #1:} }{\hfill $\blacksquare$ \\}

\let\oldref\ref
\renewcommand{\ref}[1]{(\oldref{#1})}

\newcolumntype{L}[1]{>{\raggedright\let\newline\\arraybackslash\hspace{0pt}}m{#1}}
\newcolumntype{C}[1]{>{\centering\let\newline\\arraybackslash\hspace{0pt}}m{#1}}
\newcolumntype{R}[1]{>{\raggedleft\let\newline\\arraybackslash\hspace{0pt}}m{#1}}

\begin{document}

	\begin{titlepage}
		
		\title{Data and Inequality\thanks{I would like to thank Francesc Dilmé, Stephan Lauermann, Nicola Limodio, Benny Moldovanu, Volker Nocke, Oliver Pfäuti, and Marcel Preuss for insightful comments. Funding by the Deutsche Forschungsgemeinschaft (German Research Foundation) through CRC TR 224 is gratefully acknowledged.
		}}
		\author{Carl-Christian Groh\thanks{Groh: Institute for Microeconomics, University of Bonn. Lennéstrasse 37, 53113 Bonn, Germany (email: \texttt{carlchristian.groh@gmail.com}) }} 
		\date{\today}
		\maketitle
		\begin{abstract}
			\noindent
Who receives the value generated by data? I consider a model of production in which labor
can be flexibly adjusted, and access to data enables firms to condition their capital demand
on their future profitability. Wider access to data increases the equilibrium wage if capital and labor are
complements. A monopolistic data provider restricts access if the factors are strong
complements or strong substitutes, because this raises the value of data by reducing the
scarcity of the factor firms with data use more intensively. Market power in data provision thus
harms workers if factors are strong complements.
			\\
			\\
			\textbf{Keywords:} inequality, big data, uncertainty, wages 
			\\
			\textbf{JEL Classification:} D21, D81, E24, J24 
			
			\bigskip
		\end{abstract}
		\setcounter{page}{0}
		\thispagestyle{empty}
	\end{titlepage}
	\pagebreak \newpage
	
	\section{Introduction}
 Modern economies increasingly revolve around (big) data, i.e., digitized information about
consumers and market conditions. 
By 2015, roughly three-quarters of U.S.\ manufacturing plants were already implementing some form of predictive analytics \citep{brynjolfsson2021power}. Moreover, the value generated by data is enormous \citep{farboodi2022data}. As advances in artificial intelligence unlock ever more of the predictive value of data \citep{agrawal2019artificial}, two
questions become increasingly pressing:
Who receives the economic value generated by data? Moreover, who benefits and who loses from the fact that access to data for predictive analytics is sold by providers with market power?

To answer these questions, I consider a simple theoretical model of production in
which an exogenous share of firms has access to data. Firms produce output using capital and
labor, and can flexibly hire labor on the spot market. In the absence of access to data, firms must choose their capital input under uncertainty regarding their future profitability. This
specification represents an empirically well-documented time-to-build friction: It takes firms
longer to adjust their capital stocks than their labor inputs
\citep{kydland1982time, koeva2000facts, davis2013the}. Access to data resolves the uncertainty. In particular, a firm with data learns its profitability before choosing its capital stock. This captures a central feature of data which distinguishes it from existing technologies: data enables firms to predict key payoff-relevant states such as demand, costs, or their rivals' strategic choices.

Increases in the availability of data, i.e., in the number of firms with data, benefit workers if capital and labor are complements,
and benefit capital owners if the input factors are substitutes.
Two effects determine how data availability affects the equilibrium wage and the rental
rate of capital, namely an \textit{output effect} and a \textit{relative scarcity effect}. The output
effect captures that a larger share of informed firms raises total income, which increases
both the wage and the rental rate of capital, ceteris paribus. However, the availability of
data also changes the relative scarcity of the two input factors, which is captured by the
relative scarcity effect.
When the input factors are complements, capital is the scarce
factor in the absence of data because firms with a low realized profitability hold
inefficiently large amounts of capital.
When the input factors are substitutes,
labor is the scarce factor in the absence of data because highly profitable firms hire a
correspondingly large amount of labor. Increases in the availability of
data reduce these imbalances and therefore harm the input factor that is scarce in the absence of data, i.e., capital when the
inputs are complements, and labor when the inputs are substitutes.

Further, increases in the availability of data reduce the value of data, i.e., the gap between the
profits of a firm with access to data and the profits of a firm without it. Intuitively, the
diffusion of data reduces the value of access to data by increasing the relative
scarcity of the factor that firms with data use more intensively. Importantly, this
mechanism is weak if the elasticity of substitution between the factors is intermediate, but
strong if capital and labor are strong complements or strong substitutes. In these cases, one factor is particularly scarce in the absence of data, so the diffusion
of data triggers particularly large changes in the relative scarcity of factors, and hence in the value of data.

A monopolistic data provider thus optimally sells data access to all firms if the elasticity of substitution is intermediate, but not if capital and
labor are sufficiently complementary or sufficiently
substitutable. Every firm has a positive (and identical) willingness to pay
for data, so expanding access raises the provider's profits, ceteris paribus.
However, restricting access to data increases every firm's willingness to pay for data through the aforementioned relative scarcity effect. Because this effect is particularly strong if factors are strong complements or strong substitutes, the provider optimally restricts access to data in these cases. 

If the input factors are sufficiently complementary, workers are hence worse off when there is a
monopolistic data provider than under perfect competition, where the price of data equals
its marginal cost. This is because workers benefit from the diffusion of data, but the
data provider optimally restricts access. The empirical evidence strongly suggests that capital and labor are indeed complements
\citep{oberfield2021micro,gechert2022measuring}, which indicates that market power in data provision may harm workers in practice.
\\
\\
\textbf{Related Literature:} This paper contributes to the growing literature on the
macroeconomic effects of data. To the best of my knowledge, this is the first paper to study how data, understood as firms'
forecasts of their idiosyncratic profitability, affects the incomes of labor and capital,
and how market power in the provision of data shapes these effects.

The literature on the macroeconomic effects of data, surveyed by \citet{veldkamp2024data}, has studied how data shapes growth
\citep{jones2020nonrivalry, farboodi2021model} and markups \citep{eeckhout2022data}.
\citet{asriyan2025macroeconomics} document that the expectations of U.S. firms have become
much more accurate since the early 2000s and study the consequences for productivity and
welfare. \cite{babina2024artificial} analyze the empirical relationship between AI adoption and
the within-firm workforce composition. Importantly, none of these papers studies how data affects the
distribution of income between labor and capital. Closest
in terms of focus, \citet{abis2024changing} estimate that big-data technologies lower the
labor share in investment management, but data in their paper is an input into knowledge
production, not a forecast that guides input choices.

The information friction in my model has been studied in the literature on misallocation.
\citet{david2016information} consider firms that choose inputs under imperfect information
about their fundamentals and quantify the resulting losses in aggregate productivity and
output. \citet{david2019sources} separate this friction from other sources of capital
misallocation. I take this friction as given and study how the availability of data that
resolves it affects factor incomes, and how a monopolistic data provider optimally sells
such data.

My analysis also relates to the literature on how new technologies affect wages and factor
shares \citep{zeira1998workers, karabarbounis2014global, acemoglu2018race, caselli2019robot,
	autor2020fall, acemoglu2022tasks}. In this literature, technology changes how inputs are
combined in production, for example by automating tasks. By contrast, data in my model
leaves the production technology unchanged and only improves the information that firms have
when they choose their capital.

Finally, my analysis of the data provider relates to the literature on the sale of
information, as surveyed by \cite{bergemann2019markets}. 
\citet{admati1986monopolistic} and \cite{admati1990direct} have demonstrated that a monopolist
may restrict the supply or the precision of information because buyers impose negative
externalities on each other. In \citet{kastl2018selling} and \citet{bimpikis2019information},
the closest papers, these externalities arise in the product market. In my model, they arise
due to the presence of factor markets: informed firms bid up the prices of the inputs they rely on. The provider
therefore restricts access both when capital and labor are strong complements and when they
are strong substitutes.
	
	\section{Baseline Model}
	\label{sec:model}
	\textit{Firms and production:}
	There is a unit mass of firms, each run by an entrepreneur who receives the profits of their firm.
	Workers inelastically supply one unit of labor in total and capital owners
	inelastically supply one unit of capital in total. Factor markets are competitive.
	The wage rate is denoted by $w$, and the rental rate for capital is denoted by $r$.
	
	A firm with idiosyncratic profitability $A\geq 0$ that employs $K$ units of capital
	and $N$ units of labor earns revenue given by
	\begin{equation}\label{eq:revenue}
		R(A,K,N) \;=\; A\,G(K,N)^{\theta},
	\end{equation}
	where $G(K,N)=\bigl(aK^{\gamma}+bN^{\gamma}\bigr)^{1/\gamma}$, $\theta\in(0,1)$,
	$a,b>0$, $a+b=1$, and $\gamma<1$.\footnote{Setting both factor supplies to one and
		$a+b=1$ is a normalization of units. It implies that $a$ and $b$ are the shares of
		capital and labor in factor income when all firms have access to data.}
	For $\gamma=0$, $G$ is defined by its limit, $G(K,N)=K^{a}N^{b}$. The elasticity of
	substitution between capital and labor is $\sigma=1/(1-\gamma)$. The parameter
	$\theta$ can reflect decreasing returns to scale or a downward-sloping demand curve
	that the firm takes as given. Revenue is measured in units of an outside good.
	
	Each firm's idiosyncratic profitability $A$ is drawn independently from a
	distribution $F$ on $[0,\infty)$. I assume that $F$ is non-degenerate, that its
	support contains zero, and that $\mathbb{E}\bigl[A^{1/(1-\theta)}\bigr]<\infty$.
	There is no aggregate uncertainty.
	
	\textit{Data and timing:}
	A share $\lambda\in[0,1]$ of firms has access to data. A firm with access to data
	observes its own profitability before making any decision. The remaining firms have
	no access to data and know only $F$ when choosing their capital stock. I refer to
	firms with access to data as \emph{informed} and to the remaining firms as
	\emph{uninformed}. The timing is as follows: In the first period, firms choose how
	much capital to rent. Informed firms know their profitability when making this
	choice, whereas uninformed firms do not. In the second period, uninformed firms
	also learn their profitability. All firms then choose how much labor to hire, and
	production takes place.
	
	\textit{Equilibrium:}
	Given the factor prices $(w,r)$, an informed firm chooses capital and labor knowing its
	profitability. Its expected profit is
	\begin{equation}\label{eq:pi1}
		\pi_{1}(w,r) = \mathbb{E}\Bigl[\,\max_{K,N\geq 0}\;
		\bigl \{
		A\,G(K,N)^{\theta}-wN-rK\bigr \} \Bigr].
	\end{equation}
	The expected profits of an uninformed firm are given by
	\begin{equation}\label{eq:pi0}
		\pi_{0}(w,r) = \max_{K\geq 0}\;
		\biggl \{
		\mathbb{E}\Bigl[\,\max_{N\geq 0}
		\bigl \{ A\,G(K,N)^{\theta}-wN  	\bigr \} \Bigr]-rK 	\biggr \} .
	\end{equation}
	In both expressions, the expectation is taken over the firm's profitability.
	Let $K_{1}(A)$ and $N_{1}(A)$ denote the capital and labor choices of an informed firm, and $K_{0}$
	and $N_{0}(A)$ the capital and labor choices of an uninformed firm.
	
	For given $\lambda$, an equilibrium consists of prices $(w,r)$ and input choices
	such that all firms choose their inputs optimally and both factor markets clear,
	i.e.:
	\begin{align}
		\lambda\,\mathbb{E}[K_{1}(A)]+(1-\lambda)\,K_{0} &= 1, \label{eq:Kclear}\\
		\lambda\,\mathbb{E}[N_{1}(A)]+(1-\lambda)\,\mathbb{E}[N_{0}(A)] &= 1. \label{eq:Nclear}
	\end{align}
	
	I denote the equilibrium wage by $w(\lambda, \sigma)$, the equilibrium rental rate for
	capital by $r(\lambda, \sigma)$, the expected profit of an informed firm by
	$\pi_{1}(\lambda, \sigma)$, and the expected profit of an uninformed firm by
	$\pi_{0}(\lambda, \sigma)$. Total income, that is, aggregate revenue, is denoted by
	$Y(\lambda, \sigma)$. Since revenue is homogeneous of degree $\theta$ in the two inputs,
	factor owners receive $w(\lambda, \sigma)+r(\lambda, \sigma)=\theta\,Y(\lambda, \sigma)$ and entrepreneurs
	receive $\lambda\pi_{1}(\lambda, \sigma)+(1-\lambda)\pi_{0}(\lambda, \sigma)=(1-\theta)\,Y(\lambda, \sigma)$.\footnote{One can show that the key results of the paper extend even when revenue is non-homogeneous.}

	\section{Analysis: Exogenous Data}
	
	In this section, I establish who benefits from the availability of data. I begin by characterizing how increases in the availability of data in the economy (modeled as increases in the share of informed firms) affect the wage rate and the rental rate of capital:
	
	\begin{proposition}
		[Data and Factor Income]	
		\label{prop:exog:factorincome}
		\hfill \break
		Consider the baseline model with exogenous data availability. 
		\begin{itemize}
			\item Suppose $\sigma \leq 1$. Then, the equilibrium wage is strictly increasing in $\lambda$ for all $\lambda \in [0,1]$.
			\item Suppose $\sigma > 1$. The 
			equilibrium wage is strictly higher when all firms are informed than when none are informed if and only if $\sigma < \frac{1}{1-\theta}$.
			\item If $\sigma \geq 1$, the equilibrium rental rate is strictly increasing in $\lambda$. Further,  $\lim_{ \sigma \rightarrow 0} r(0, \sigma) \leq \lim_{ \sigma \rightarrow 0} r(1, \sigma)$ if and only if 
			$\mathbb{E}[A] \leq a \big(\mathbb{E}[A^{1/(1-\theta) } ]\big)^{(1-\theta)}$.
		\end{itemize}
	\end{proposition}
	This proposition formalizes the following results. First, when capital and labor are
	complements ($\sigma \leq 1$), an increase in the share of informed firms raises the
	equilibrium wage for any distribution of profitability. Second, when capital and labor
	are substitutes ($\sigma > 1$), the wage need not be monotone in the share of informed
	firms, and its path depends on the spread of the profitability distribution. Nevertheless,
	the wage is higher when all firms are informed than when none are informed as long as
	capital and labor are not too easily substitutable, that is, as long as
	$\sigma < \frac{1}{1-\theta}$. Third, if capital and labor are substitutes
	($\sigma \geq 1$), the equilibrium rental rate of capital is strictly increasing in data
	availability, again for any distribution of profitability. If they are complements, the effect is more nuanced and generally
	non-monotonic. For example, when capital and labor are close to perfect complements, the
	rental rate is higher when all firms are informed than when none are informed if and only
	if capital's share of factor income and the spread of profitability are sufficiently large.
	
	Intuitively, two effects determine how data availability affects the wage and the rental
	rate, namely an \textit{output effect} and a \textit{relative scarcity effect}. The output
	effect captures that a larger share of informed firms raises total income. Because revenue
	is homogeneous of degree $\theta$ in capital and labor, factor owners receive a constant
	share of total income, so total factor payments rise as well. Ceteris paribus, this
	benefits both workers and capital owners.
	
	The relative scarcity effect captures how the availability of data changes the relative scarcity of each input factor. 
	When capital and
	labor are highly complementary, i.e., when $\sigma$ is small, capital is the scarce
	factor if firms are uninformed. The reason is that all firms without data hold the same
	amount of capital because they cannot condition this on their profitability, so capital is in short supply at the most profitable firms. Increasing
	the share of informed firms reduces this scarcity, because firms with low profitability no
	longer hold significant amounts of capital. By itself, this lowers the rental rate of
	capital and raises the wage. The relative scarcity effect goes in the opposite direction when capital and labor are highly
	substitutable, i.e., when $\sigma$ is large. Then, highly profitable uninformed firms demand a large amount of labor, since labor is the only input they can match to their profitability. Data availability removes this advantage of labor, which
	makes labor less scarce relative to capital. In sum, the relative scarcity effect benefits
	workers and harms capital owners if the inputs are highly complementary, and vice versa if they are highly substitutable.
	
	Taken together, these effects imply that capital owners benefit and workers are harmed
	by data availability if the inputs are highly substitutable.  If the inputs are
 complementary, workers benefit from greater data availability. Capital owners benefit
	as well if the output effect on their income is large enough to outweigh the relative
	scarcity effect, which is the case if their share of factor income and the spread of
	profitability are large.

	I now characterize how the availability of data affects profits and the value of information, i.e., the difference between the profits of an informed and an uninformed firm.
	
	\begin{proposition}[Data and Firm Profits]
		\label{prop:exog:profits}
		\hfill \break
		Consider the baseline model with exogenous data availability. Total profits are strictly
		increasing in $\lambda$. Further, 
		the gap $\pi_{1}(\lambda)-\pi_{0}(\lambda)$ is positive and strictly decreasing in
		$\lambda$. 
	\end{proposition}
	\noindent
	The first result in the proposition is that total profits rise with the share of informed
	firms. Wider access to data raises total output, and entrepreneurs receive a
	fixed share $1-\theta$ of it because revenue is homogeneous of degree $\theta$. The second
	result in the proposition is that improved access to data reduces the value of information, that is, the gap
	between the profits of an informed and an uninformed firm. Intuitively, a larger share of
	informed firms raises the cost of the input factors that firms use, which lowers
	the profits of any firm, ceteris paribus.
	An uninformed firm is hurt less, and may even gain, for two reasons. First, it spends less
	on inputs than an informed firm. Second, relative factor prices shift against the input
	that informed firms use more intensively, which is capital if $\sigma>1$ and labor if
	$\sigma<1$.

	\section{Optimal Selling of Data}
	
	\subsection{Model}
	\label{sec:model:sales}
	\noindent
	I now consider a version of the baseline model in which the availability of data is endogenous because data is sold by a monopolistic data provider. Formally, I augment the model from Section \ref{sec:model} with an additional initial stage: In the first period, a data provider now posts a price $p\geq 0$ at
	which any firm can buy access to data. Firms observe this price and, without knowing their profitability, decide whether to purchase data. If they purchase data, they become informed, and they remain uninformed otherwise. 
	Thereafter, the game proceeds as in Section
	\ref{sec:model}, where $\lambda$ is the share of firms that have purchased data. The data provider can serve any number of firms at zero
	cost.
	
	A firm is willing to pay at most the gap between the expected profits of an informed and
	an uninformed firm for access to data. I denote this gap by
	\begin{equation}\label{eq:seller:gap}
		D(\lambda) \;\equiv\; \pi_{1}(\lambda)-\pi_{0}(\lambda).
	\end{equation}
	The equilibrium concept is subgame-perfect equilibrium. An equilibrium consists of a
	price $p^{*}$ and a function $\hat{\lambda}(p)$ with the following properties. For any
	$p$, the share $\hat{\lambda}(p)$ of firms that purchase data must be such that no firm
	would want to deviate from its choice: If $\hat{\lambda}(p) = 1$ (respectively,
	$\hat{\lambda}(p) = 0$), all firms must weakly prefer to purchase data (respectively,
	not purchase data), given that all other firms purchase data (respectively, do not
	purchase data). If $\hat{\lambda}(p) \in (0,1)$, all firms must be indifferent between
	purchasing data and not. The price $p^{*}$ must maximize the profits of the data provider,
	namely $p\,\hat{\lambda}(p)$. I refer to the share of firms that purchase data in equilibrium as $\lambda^* = \hat{\lambda}(p^*)$.
	
	\subsection{Analysis}
	\noindent
	I directly characterize the optimal policy of the data provider. 
	
	\begin{proposition}[Optimal Sale of Data]
		\label{prop:endog:optimal}
		\hfill \break
		Consider the model with a monopolistic data provider.
		\begin{itemize}
			\item Suppose $1\leq\sigma\leq\frac{1}{1-\theta}$. Then, the data provider sells data
			to all firms, i.e., $\lambda^{*}=1$.
			\item There exist $\sigma_{L}\in(0,1)$ and $\sigma_{H}>\frac{1}{1-\theta}$
			such that $\lambda^{*}<1$  if $\sigma<\sigma_{L}$ or $\sigma>\sigma_{H}$.
		\end{itemize}
	\end{proposition}
	\noindent
	This proposition formalizes that the data provider restricts access to data if capital and
	labor are sufficiently complementary (i.e., if $\sigma$ is small) or sufficiently
	substitutable (i.e., if $\sigma$ is large), but does not restrict access if the elasticity of substitution is intermediate. Intuitively, two effects determine the optimal
	selling policy. On the one hand, every firm has a strictly positive willingness to pay
	for data, so expanding access raises the profits of the data provider, ceteris paribus.
	On the other hand, restricting access to data raises every firm's willingness to pay for data. This
	is because restricting access to data 
	raises the scarcity of the input that uninformed firms rely on more heavily than informed firms.  
	As formalized in the previous section, 
	this relative
	scarcity effect is strong if $\sigma$ is far from one. The data
	provider thus restricts access to data if $\sigma$ is sufficiently low or
	high.
	
	Taking these results into account, I now establish under what conditions workers and
	capital owners benefit from or are hurt by the market power of the data provider. To that
	end, I compare the outcome under the optimal pricing strategy of the monopolistic data
	provider to the outcome under perfect competition among data providers, i.e., when the price of data is equal to the marginal cost of data provision, namely zero. 
	
	\begin{corollary}[Market Power and Factor Incomes]
		\label{prop:seller:incomes}
		\hfill \break
		If $\sigma < \sigma_L$, the equilibrium wage is strictly lower if there is a single data
		provider than under perfect competition, and the rental rate is strictly higher. If
		$\sigma > \sigma_H$, the equilibrium wage is strictly higher if there is a single data
		provider than under perfect competition, and the rental rate is strictly lower. If
		$\sigma \in \big[1, \frac{1}{1-\theta}\big]$, each factor price is the same in both
		cases.
	\end{corollary}
	\noindent
	Thus, market power in the sale of data hurts workers and benefits capital owners if
	capital and labor are strongly complementary. The reason is that, if capital and labor
	are strong complements, workers benefit if the share of informed firms increases, given
	that this increases the relative scarcity of labor (Proposition
	\ref{prop:exog:factorincome}). However, the monopolistic data provider restricts access to
	data (Proposition \ref{prop:endog:optimal}). When capital and labor are strongly
	substitutable, this is reversed. If the elasticity of substitution is intermediate, the data
	provider does not restrict access to data, so the outcome is identical to the outcome
	under perfect competition.
	
	\section{Extensions}

	\textbf{Noisy information:} In Section \ref{sec:supmat:noisy} of the Online Appendix, I consider an extension of the baseline model in which firms' information is noisy. Specifically, profitability is log-normally distributed, and all firms exogenously have access to a noisy signal about their future profitability, on which they can condition their capital choice. I prove that an increase in the precision of the signal has qualitatively analogous effects on the equilibrium wage, the rental rate, total firm profits, and the value of information as an increase in the share of informed firms in the baseline analysis.
	Intuitively, the time-to-build friction forces firms to adjust labor ex post to compensate for suboptimal capital choices. Whether this mismatch is eliminated entirely for more firms (via a larger share of informed firms) or reduced slightly for every firm (via more precise signals), capital is allocated more efficiently. The effect of this on the relative scarcity of the factors, and therefore on their remuneration, is pinned down by the elasticity of substitution, as before.
	\\
	\\
	\textbf{Elastic factor supplies:} In Section \ref{sec:supmat:elastic} of the Online Appendix, I consider an extension of the baseline model in which factor supplies are elastic. An exogenous share of firms has access to perfect information about their future profitability, while the other firms know nothing beyond the prior distribution. I prove that, as in the baseline analysis, an increase in the share of informed firms raises the equilibrium wage if the factors are complements, raises the rental rate if they are substitutes, and reduces the value of information. The effect of an increase in the share of informed firms on total profits is more nuanced and depends on the factor supply elasticities: total profits increase if factors are supplied equally elastically, but can potentially decrease if data shifts demand toward the more inelastically supplied factor.

	\section{Conclusion}
	
Modern economies increasingly revolve around (big) data, i.e., digitized information which
enables firms to forecast payoff-relevant states. I study the distributional consequences of
these developments in a simple model of production.  The equilibrium wage
increases when access to data diffuses if capital and labor are complements. Furthermore, the diffusion of data reduces the value of access to data by increasing the relative scarcity of the factor that firms with data use more
intensively.
A monopolistic data provider thus optimally sells data to all firms if the elasticity of substitution
between the factors is intermediate, but restricts access if the factors are strong
complements or strong substitutes because the relative scarcity effect dominates. In the empirically most relevant case where capital and labor are strong complements in production, market power in data provision can thus harm workers.

	\appendix
	\section*{Proofs}

I first establish three helpful Lemmas. 

\begin{lemma}\label{lem:duality}
	Let $\lambda\in[0,1]$, and let $K_{1}(A)$ and $N_{1}(A)$ denote the inputs of an informed
	firm and $K_{0}$ and $N_{0}(A)$ those of an uninformed firm. Input choices
	$(K_{1}(A),N_{1}(A),K_{0},N_{0}(A))$ are \emph{feasible at $\lambda$} if they satisfy
	\eqref{eq:Kclear} and \eqref{eq:Nclear} with weak inequalities. Their aggregate revenue
	under a production function $\tilde{G}$ is
	$\lambda\,\mathbb{E}[A\,\tilde{G}(K_{1}(A),N_{1}(A))^{\theta}]
	+(1-\lambda)\,\mathbb{E}[A\,\tilde{G}(K_{0},N_{0}(A))^{\theta}]$, and the profit functions
	$\tilde{\pi}_{1}$ and $\tilde{\pi}_{0}$ of $\tilde{G}$ are defined as in \eqref{eq:pi1} and
	\eqref{eq:pi0}, with $\tilde{G}$ in place of $G$.
	\begin{itemize}
		\item[(a)] For any production function $\tilde{G}$, any input choices that are feasible at
		$\lambda$, and any prices $w,r\geq0$, aggregate revenue under $\tilde{G}$ is at most
		\begin{equation}\label{eq:weakdual}
			\lambda\,\tilde{\pi}_{1}(w,r)+(1-\lambda)\,\tilde{\pi}_{0}(w,r)+w+r.
		\end{equation}
		\item[(b)] In any equilibrium, aggregate revenue $Y(\lambda)$ satisfies
		\begin{equation}\label{eq:dual}
			Y(\lambda)=\min_{w,r>0}\bigl\{\lambda\pi_{1}(w,r)+(1-\lambda)\pi_{0}(w,r)+w+r\bigr\},
		\end{equation}
		the equilibrium prices attain the minimum, and $Y(\lambda)$ is the highest aggregate
		revenue among all input choices that are feasible at $\lambda$. Moreover, for all
		$\lambda,\lambda'\in[0,1]$,
		\begin{equation}\label{eq:supergradient}
			Y(\lambda')\leq Y(\lambda)+(\lambda'-\lambda)\bigl[\pi_{1}(\lambda)-\pi_{0}(\lambda)\bigr].
		\end{equation}
	\end{itemize}
\end{lemma}

\noindent
\begin{App-Proof}{Lemma \ref{lem:duality}}
	(a) By the definition of the profit functions, the expected revenue of an informed firm
	is at most $\tilde{\pi}_{1}(w,r)+w\,\mathbb{E}[N_{1}(A)]+r\,\mathbb{E}[K_{1}(A)]$, and
	that of an uninformed firm is at most
	$\tilde{\pi}_{0}(w,r)+w\,\mathbb{E}[N_{0}(A)]+r\,K_{0}$. Taking the weighted sum over
	both types of firms and using feasibility yields the claim. (b) In an equilibrium, all
	input choices are optimal and both factor markets clear, so the aggregate revenue
	$Y(\lambda)$ equals \eqref{eq:weakdual} for $\tilde{G}=G$ at the equilibrium prices.
	Together with (a), this proves \eqref{eq:dual} and the two claims that follow it.
	Applying (a) to the equilibrium input choices for $\lambda'$ and the equilibrium prices
	for $\lambda$ yields \eqref{eq:supergradient}.
\end{App-Proof}

\begin{lemma}\label{lem:equilibrium}
	Throughout, let $m\equiv 1/(1-\theta)$, $t\equiv\theta/(1-\theta)$ and $\rho\equiv r/w$.
	The following properties hold, the first two at any prices $w,r>0$.
	\begin{itemize}
		\item[(P1)] \emph{First-order conditions.} A firm with $A>0$ chooses $N$ such that
		\begin{equation}\label{eq:focN}
			\theta b\,A\,N^{\gamma-1}G(K,N)^{\theta-\gamma}=w,
		\end{equation}
		and a firm with $A=0$ hires no labor. An informed firm with $A>0$ chooses $K$ such
		that $\theta a\,A\,K^{\gamma-1}G(K,N)^{\theta-\gamma}=r$, whereas an uninformed firm
		chooses $K_{0}$ such that
		\begin{equation}\label{eq:focK0}
			\theta a\,K_{0}^{\gamma-1}\,\mathbb{E}\bigl[A\,G(K_{0},N_{0}(A))^{\theta-\gamma}\bigr]=r.
		\end{equation}
		Moreover, $N_{0}(A)$ is strictly increasing in $A$.
		\item[(P2)] \emph{Factor payments and homogeneity.} The expected factor payments of
		each firm equal $\theta$ times its expected revenue, and hence $t$ times its expected
		profit. Moreover, profits are homogeneous of degree $-t$ and input choices are
		homogeneous of degree $-m$ in $(w,r)$.
		\item[(P3)] \emph{Bounds for complements.} Let $\sigma<1$ and
		$\kappa_{\sigma}\equiv(\min\{a,b\})^{-\sigma/(1-\sigma)}$, so that
		$\kappa_{\sigma}\to1$ as $\sigma\to0$. Then
		\begin{equation}\label{eq:sandwich}
			\min\{K,N\}\leq G(K,N)\leq\kappa_{\sigma}\min\{K,N\}.
		\end{equation}
	\end{itemize}
	Moreover, for every $\lambda\in[0,1]$, there exists a unique equilibrium. For $\lambda>0$,
	the equilibrium prices are the unique minimizer in \eqref{eq:dual}. Finally,
	$w(\lambda)$, $r(\lambda)$, $\pi_{1}(\lambda)$, $\pi_{0}(\lambda)$ and $Y(\lambda)$ are
	continuous in $\lambda$. I write $w^*(\lambda,\sigma)$ and $r^*(\lambda,\sigma)$, or add
	the subscript $\sigma$, when the dependence on $\sigma$ matters.
\end{lemma}

\noindent
Lemma 2 establishes that the problem is well-defined. 
Its proof can be found in Section \ref{sec:supmat:equilibrium} of the
Online Appendix.

\begin{lemma}\label{lem:intensity}
	Fix prices $w,r>0$. Let $K_{1}$, $N_{1}$ and $N_{0}$ denote the expectations of
	$K_{1}(A)$, $N_{1}(A)$ and $N_{0}(A)$, and let $s_{1}$ and $s_{0}$ denote the shares of
	capital in the expected factor payments of an informed and an uninformed firm. Then
	$N_{1}(A)/K_{1}(A)=N_{1}/K_{1}$ for all $A>0$, and
	\begin{equation}\label{eq:intensity}
		\Bigl(\frac{N_{1}}{K_{1}}\Bigr)^{1/\sigma}
		=\frac{b}{a}\,\rho
		=\mathbb{E}\Bigl[\Bigl(\frac{N_{0}(A)}{K_{0}}\Bigr)^{1/\sigma}\Bigr].
	\end{equation}
	Hence $N_{1}/K_{1}<N_{0}/K_{0}$ if $\sigma>1$, equality holds if $\sigma=1$, and the
	reverse inequality holds if $\sigma<1$. Accordingly, $s_{1}>s_{0}$ if $\sigma>1$,
	$s_{1}=s_{0}$ if $\sigma=1$, and $s_{1}<s_{0}$ if $\sigma<1$.
\end{lemma}

\noindent
\begin{App-Proof}{Lemma \ref{lem:intensity}}
	Dividing the two first-order conditions of an informed firm in (P1) yields
	$(a/b)(N/K)^{1/\sigma}=\rho$, so $N_{1}(A)/K_{1}(A)=N_{1}/K_{1}$ for all $A>0$. For an
	uninformed firm, \eqref{eq:focN} implies
	$\theta A\,G^{\theta-\gamma}=wN_{0}(A)^{1/\sigma}/b$. Substituting this into
	\eqref{eq:focK0} yields \eqref{eq:intensity}. By (P1), $N_{0}(A)$ is non-degenerate, so
	Jensen's inequality implies the stated ranking of $N_{1}/K_{1}$ and $N_{0}/K_{0}$, and
	hence that of $s_{1}$ and $s_{0}$.
\end{App-Proof}

\noindent
\begin{App-Proof}{Proposition \ref{prop:exog:factorincome}}
	\par\medskip
	\noindent\textbf{Part 1:} Preliminaries.
	
	\medskip
	\noindent\textit{(i) Total income and factor prices.}
	By (P2) and market clearing, $w(\lambda)+r(\lambda)=\theta\,Y(\lambda)$, so that
	\begin{equation}\label{eq:shares}
		w(\lambda)=\frac{\theta\,Y(\lambda)}{1+\rho(\lambda)}
		\qquad\text{and}\qquad
		r(\lambda)=\frac{\theta\,Y(\lambda)\,\rho(\lambda)}{1+\rho(\lambda)}.
	\end{equation}
	Moreover, $Y(\lambda)$ is strictly increasing in $\lambda$: for $\lambda<\lambda'$,
	\eqref{eq:supergradient} with the roles of $\lambda$ and $\lambda'$ exchanged yields
	$Y(\lambda')-Y(\lambda)\geq(\lambda'-\lambda)[\pi_{1}(\lambda')-\pi_{0}(\lambda')]>0$, because an
	informed firm can always choose $K_{0}$ and its optimal capital stock varies with $A$.
	
	\medskip
	\noindent\textit{(ii) The relative factor price.}
	I show that $\rho(\lambda)$ is strictly increasing in $\lambda$ if $\sigma>1$, constant if
	$\sigma=1$, and strictly decreasing if $\sigma<1$. By (P2),
	$\pi_{j}(w,r)=w^{-t}g_{j}(\rho)$ with $g_{j}(\rho)\equiv\pi_{j}(1,\rho)$. Let
	$M(\rho,\lambda)\equiv\lambda g_{1}(\rho)+(1-\lambda)g_{0}(\rho)$. By Lemma \ref{lem:duality},
	the equilibrium prices minimize $w^{-t}M(\rho,\lambda)+w(1+\rho)$. Minimizing over $w$ for given
	$\rho$ shows that $\rho(\lambda)$ minimizes
	\begin{equation*}
		\Psi(\rho,\lambda)\equiv(1-\theta)\ln M(\rho,\lambda)+\theta\ln(1+\rho).
	\end{equation*}
	By Hotelling's lemma and (P2), $\rho\,g_{j}'(\rho)=-t\,s_{j}\,g_{j}(\rho)$. Therefore,
	\begin{equation*}
		\frac{\partial^{2}\Psi}{\partial\rho\,\partial\lambda}
		=(1-\theta)\,\frac{g_{1}'g_{0}-g_{0}'g_{1}}{M^{2}}
		=\frac{\theta\,g_{0}\,g_{1}\,(s_{0}-s_{1})}{\rho\,M^{2}},
	\end{equation*}
	which by Lemma \ref{lem:intensity} is negative if $\sigma>1$, zero if $\sigma=1$, and positive if $\sigma<1$. If
	$\sigma>1$, $\Psi$ thus has strictly decreasing differences in $(\rho,\lambda)$, so its minimizer
	$\rho(\lambda)$ is weakly increasing in $\lambda$. It is strictly increasing because the
	first-order condition $\partial\Psi/\partial\rho=0$ cannot hold at the same $\rho$ for two
	different values of $\lambda$. The case $\sigma<1$ is analogous. If $\sigma=1$, the first-order
	condition does not depend on $\lambda$.
	
	\medskip
	\noindent\textit{(iii) The equilibrium when all firms are informed.}
	Let $\lambda=1$. By Lemma \ref{lem:intensity} and market clearing, $N/K=1$ at all firms, so \eqref{eq:intensity}
	implies $\rho(1)=a/b$. Moreover, $G(K,N)=N$, so \eqref{eq:focN} yields
	$N_{1}(A)=(\theta bA/w)^{m}$. Let $Y_{F}\equiv(\mathbb{E}[A^{m}])^{1/m}$. Labor market
	clearing and (i) then yield
	\begin{equation}\label{eq:fullinfo}
		w(1)=\theta\,b\,Y_{F},
		\qquad
		r(1)=\theta\,a\,Y_{F}
		\qquad\text{and}\qquad
		Y(1)=Y_{F}.
	\end{equation}
	
	\medskip
	\noindent\textit{(iv) A representation of the wage.}
	Let $\gamma\neq 0$, $\tilde g(y)\equiv(ay^{\gamma}+b)^{1/\gamma}$ and
	$\psi(y)\equiv\tilde g(y)^{(\gamma-\theta)m}$. Since $G(K,N)=N\,\tilde g(K/N)$, raising
	\eqref{eq:focN} to the power $m$ yields $A^{m}=(w/(\theta b))^{m}\,N\,\psi(K/N)$ for every firm
	in every state, where the right-hand side is zero if $A=0$. Taking the expectation over all
	firms and states for $\lambda=0$, where every firm holds $K_{0}=1$, and using
	\eqref{eq:fullinfo} yields
	\begin{equation}\label{eq:endpoints}
		\Bigl(\frac{w(1)}{w(0)}\Bigr)^{m}=\mathbb{E}\bigl[\varphi(n(A))\bigr],
	\end{equation}
	where $n(A)$ is the labor of a firm with profitability $A$ in the absence of data,
	$\mathbb{E}[n(A)]=1$, $\varphi(n)\equiv n\,\psi(1/n)$ for $n>0$ and $\varphi(0)\equiv0$.
	
	\medskip
	\noindent\textit{(v) Strong complements.}
	I show that $\lim_{\sigma\to0}r^*(0,\sigma)=\theta\,\mathbb{E}[A]$. Let $\lambda=0$ and
	$\sigma\leq\frac12$, and define
	$\bar c\equiv(\min\{a,b\})^{-2}\geq\kappa_{\sigma}^{1/\sigma}$. By Lemma
	\ref{lem:duality}, the equilibrium allocation of labor maximizes
	$\mathbb{E}[A\,G(1,n(A))^{\theta}]$ subject to
	$\mathbb{E}[n(A)]=1$. Since $n(A)=1$ is feasible, \eqref{eq:sandwich} implies
	$\mathbb{E}[A]\leq Y_{\sigma}(0)\leq\kappa_{\sigma}^{\theta}\,\mathbb{E}[A]$, so
	$Y_{\sigma}(0)\to\mathbb{E}[A]$ as
	$\sigma\to0$. Since $r=\theta Y-w$ by (i), it remains to show that $w^*(0,\sigma)\to0$. The
	marginal revenue of labor is
	$M_{N}(A,n)\equiv\theta b\,A\,G(1,n)^{\theta-1}(G(1,n)/n)^{1/\sigma}$, and \eqref{eq:sandwich}
	implies
	\begin{equation}\label{eq:Mbounds}
		M_{N}(A,n)\leq\theta b\,A\,(\kappa_{\sigma}/n)^{1/\sigma}\ \text{ if }n\geq1
		\qquad\text{and}\qquad
		M_{N}(A,n)\leq\theta b\,\bar c\,A\,n^{\theta-1}\ \text{ if }n\leq1.
	\end{equation}
	Suppose that $w^*(0,\sigma_{j})\geq\varepsilon>0$ along a sequence $\sigma_{j}\to0$. Since
	$M_{N}(A,n(A))=w$, the first bound implies
	$n(A)\leq\bar n_{\sigma}(A)\equiv\max\{1,\kappa_{\sigma}(\theta bA/\varepsilon)^{\sigma}\}$. For
	$A<A_{\varepsilon}\equiv\varepsilon\,2^{\theta-1}/(\theta b\,\bar c)$, both bounds imply
	$M_{N}(A,n)<\varepsilon$ for all $n\geq\frac12$, so $n(A)<\frac12$. Hence
	\begin{equation*}
		1=\mathbb{E}[n(A)]
		\leq\tfrac12\Pr(A<A_{\varepsilon})
		+\mathbb{E}\bigl[\bar n_{\sigma_{j}}(A)\,\mathbf{1}\{A\geq A_{\varepsilon}\}\bigr]
		\;\longrightarrow\;1-\tfrac12\Pr(A<A_{\varepsilon})
	\end{equation*}
	by dominated convergence, since $\bar n_{\sigma}(A)\to1$ and
	$\bar n_{\sigma}(A)\leq\bar c\,(1+\theta bA/\varepsilon)$ for all $\sigma\leq\frac12$, given
	that $\kappa_{\sigma}\leq\kappa_{\sigma}^{1/\sigma}\leq\bar c$. The limit is strictly below one
	because the support of $F$ contains zero, a contradiction.
	
	\bigskip
	\noindent\textbf{Part 2:} The wage for $\sigma\leq1$.
	By (i), $w(\lambda)=\theta\,Y(\lambda)/(1+\rho(\lambda))$ and $Y(\lambda)$ is strictly
	increasing. By (ii), $\rho(\lambda)$ is weakly decreasing if $\sigma\leq1$. Hence $w(\lambda)$ is
	strictly increasing.
	
	\bigskip
	\noindent\textbf{Part 3:} The wage for $\sigma>1$.
	Let $\gamma\in(0,1)$. Then $\tilde g$ is strictly increasing and strictly concave, $\varphi$ is
	continuous with $\varphi(1)=1$, and $\varphi''(n)=\psi''(1/n)/n^{3}$. If
	$\sigma<\frac{1}{1-\theta}$, that is, $\gamma<\theta$, then $\psi$ is a strictly decreasing and
	strictly convex transformation of $\tilde g$, so $\psi$ and $\varphi$ are strictly convex. Since
	$n(A)$ is non-degenerate, Jensen's inequality and \eqref{eq:endpoints} yield $w(1)>w(0)$. If
	$\gamma=\theta$, then $\psi\equiv1$ and $w(1)=w(0)$. If $\gamma>\theta$, then $\psi$ is a strictly
	increasing and strictly concave transformation of $\tilde g$, so $\varphi$ is strictly concave
	and $w(1)<w(0)$.
	
	\bigskip
	\noindent\textbf{Part 4:} The rental rate.
	By (i), $r(\lambda)=\theta\,Y(\lambda)\,\rho(\lambda)/(1+\rho(\lambda))$ and $Y(\lambda)$ is
	strictly increasing. By (ii), $\rho(\lambda)$ is weakly increasing if $\sigma\geq1$. Hence
	$r(\lambda)$ is strictly increasing if $\sigma\geq1$. For the second claim, (v) and
	\eqref{eq:fullinfo} yield $\lim_{\sigma\to0}r^*(0,\sigma)=\theta\,\mathbb{E}[A]$ and
	$r^*(1,\sigma)=\theta\,a\,Y_{F}$ for all $\sigma$, which proves the claim because
	$Y_{F}=(\mathbb{E}[A^{1/(1-\theta)}])^{1-\theta}$.
\end{App-Proof}

\noindent
\begin{App-Proof}{Proposition \ref{prop:exog:profits}}
	Let $D(\lambda)\equiv\pi_{1}(\lambda)-\pi_{0}(\lambda)$, and let $\lambda<\lambda'$.
	
	\textit{Profit gap.} An informed firm can always choose the capital stock of an uninformed
	firm, and its optimal capital stock varies with $A$, so $D(\lambda)>0$. Applying
	\eqref{eq:supergradient} twice, with the roles of $\lambda$ and $\lambda'$ exchanged, and
	adding yields $(\lambda'-\lambda)[D(\lambda)-D(\lambda')]\geq0$, so $D(\lambda')\leq D(\lambda)$.
	Suppose $D(\lambda')=D(\lambda)$. Then both applications of \eqref{eq:supergradient} hold with
	equality, so the equilibrium prices for $\lambda$ also attain the minimum in \eqref{eq:dual}
	for $\lambda'$. Since $\lambda'>0$, this minimum is attained only at the equilibrium
	prices for $\lambda'$ (Lemma \ref{lem:equilibrium}), so prices are the same for $\lambda$ and
	$\lambda'$. Both factor markets can clear at the same prices for two different shares of
	informed firms only if informed and uninformed firms demand the same amounts of both inputs.
	By (P2), their factor payments, and hence their profits, would then coincide, which
	contradicts $D(\lambda)>0$. Hence $D(\lambda')<D(\lambda)$.
	
	\textit{Total profits.} By (P2), total profits equal $(1-\theta)Y(\lambda)$, and
	$Y(\lambda)$ is strictly increasing in $\lambda$ by Part 1 (i) of the proof of Proposition
	\ref{prop:exog:factorincome}.
\end{App-Proof}

\noindent
\begin{App-Proof}{Proposition \ref{prop:endog:optimal}}
	Let $V(\lambda)\equiv\lambda D(\lambda)$, and recall from \eqref{eq:fullinfo} that
	$Y(1)=Y_{F}$, $w(1)=\theta bY_{F}$ and $r(1)=\theta aY_{F}$ for every $\sigma$.
	
	\bigskip
	\noindent\textbf{Part 1:} Preliminaries.
	
	\medskip
	\noindent\textit{(i) The problem of the data provider.}
	By Proposition \ref{prop:exog:profits} and Lemma \ref{lem:equilibrium}, $D$ is strictly
	positive, strictly decreasing and continuous. Hence $\hat{\lambda}(p)$ is uniquely
	determined for every $p\geq0$, and for every $\lambda\in(0,1]$, the price $p=D(\lambda)$
	implies $\hat{\lambda}(p)=\lambda$ and yields the profit $V(\lambda)$. Conversely, any
	price $p$ yields at most $V(\hat{\lambda}(p))$, because $p=D(\hat{\lambda}(p))$ if
	$\hat{\lambda}(p)\in(0,1)$, $p\leq D(1)$ if $\hat{\lambda}(p)=1$, and the profit is zero
	if $\hat{\lambda}(p)=0$. Since $V$ is
	continuous, an equilibrium exists, and in any equilibrium, $\lambda^{*}$ maximizes $V$
	on $[0,1]$.
	
	\medskip
	\noindent\textit{(ii) Two properties of $V$.}
	Since total profits equal $(1-\theta)Y(\lambda)$,
	\begin{equation}\label{eq:opt:V}
		V(\lambda)=(1-\theta)\,Y(\lambda)-\pi_{0}(\lambda).
	\end{equation}
	Moreover, \eqref{eq:supergradient} implies that for all $\lambda<\lambda'$,
	\begin{equation}\label{eq:opt:chord}
		D(\lambda)\;\geq\;\frac{Y(\lambda')-Y(\lambda)}{\lambda'-\lambda}\;\geq\;D(\lambda').
	\end{equation}
	
	\bigskip
	\noindent\textbf{Part 2: } The case $1\leq\sigma\leq\frac{1}{1-\theta}$.
	\\
	\\
	Fix $\lambda<1$. I first show that $w(\lambda)\leq w(1)$. By \eqref{eq:focN}, every firm
	satisfies
	$N=(\theta b/w)^{\frac{1}{1-\gamma}}A^{\frac{1}{1-\gamma}}G(K,N)^{\alpha}$ in every
	state, where $\alpha\equiv\frac{\theta-\gamma}{1-\gamma}\in[0,1)$ because
	$\gamma\leq\theta$. Let $\mathbb{E}_{\lambda}$ denote the average over all firms and
	states, so that $\mathbb{E}_{\lambda}[N]=\mathbb{E}_{\lambda}[K]=1$ by market clearing.
	H\"older's inequality with exponents $\frac{1-\gamma}{1-\theta}$ and $\frac{1}{\alpha}$
	(for $\alpha=0$, the first inequality below is an equality) yields
	\begin{equation*}
		1=\mathbb{E}_{\lambda}[N]
		\leq\Bigl(\frac{\theta b}{w}\Bigr)^{\frac{1}{1-\gamma}}
		\bigl(\mathbb{E}[A^{m}]\bigr)^{\frac{1-\theta}{1-\gamma}}
		\bigl(\mathbb{E}_{\lambda}[G(K,N)]\bigr)^{\alpha}
		\leq\Bigl(\frac{\theta b}{w}\Bigr)^{\frac{1}{1-\gamma}}
		\bigl(\mathbb{E}[A^{m}]\bigr)^{\frac{1-\theta}{1-\gamma}},
	\end{equation*}
	where the second inequality holds because $G$ is concave and homogeneous of degree one,
	so that $\mathbb{E}_{\lambda}[G(K,N)]\leq G(1,1)=1$. Hence $w(\lambda)\leq\theta bY_{F}=w(1)$.
	Moreover, $r(\lambda)<r(1)$ because $\sigma\geq1$ (Part 4 of the proof of Proposition
	\ref{prop:exog:factorincome}). Since $\pi_{0}(w,r)$ is decreasing in both prices,
	$\pi_{0}(\lambda)\geq\pi_{0}(1)$. Since $Y$ is strictly increasing, \eqref{eq:opt:V}
	yields $V(\lambda)<V(1)$. By (i), $\lambda^{*}=1$.
	
	\bigskip
	\noindent\textbf{Part 3:} Existence of $\sigma_{H}$.
	\\
	\\
	Let $\sigma>1$, $\lambda_{1}\equiv\frac{a}{2}$ and $\lambda_{2}\equiv\frac{1+a}{2}$, and
	let $\bar{Y}_{\infty}(\lambda)$ denote the supremum of aggregate revenue over all input
	choices that are feasible at $\lambda$ if the production function is
	$G_{\infty}(K,N)\equiv aK+bN$.
	
	\medskip
	\noindent\textit{(a) $\bar{Y}_{\infty}(\lambda_{1})<Y_{F}$.}
	Let $P\equiv\theta Y_{F}$ and consider the prices $w=bP$ and $r=aP-\eta$ with
	$\eta\in(0,aP)$. Under $G_{\infty}$, a unit of $aK+bN$ costs $P$ if it is obtained
	through labor and $P-\eta/a$ if it is obtained through capital. Informed firms thus only
	use capital, and their expected capital
	$k_{1}(\eta)=\frac{1}{a}(\theta/(P-\eta/a))^{m}\,\mathbb{E}[A^{m}]$ converges to
	$\frac{1}{a}$ as $\eta\to0$. An uninformed firm with capital $K$
	hires labor if and only if $\theta A(aK)^{\theta-1}>P$, so the derivative of its expected
	profit with respect to $z\equiv aK$ is $\eta/a-L(z)$ with
	$L(z)\equiv\mathbb{E}[\max\{P-\theta Az^{\theta-1},0\}]$. The function $L$ is
	nondecreasing, and $L(z)>0$ for all $z>0$ because the support of $F$ contains zero.
	Since a strictly positive optimal choice satisfies $L(z)=\eta/a$, the capital
	$k_{0}(\eta)$ of an uninformed firm converges to zero as $\eta\to0$.
	At the prices $(bP,aP)$, an informed firm earns $(1-\theta)Y_{F}$, an uninformed firm
	earns weakly less, and the input choices of a firm of type $j$ described above cost
	$\eta\,k_{j}(\eta)$ more. Hence
	$\tilde{\pi}_{j}(bP,aP-\eta)\leq(1-\theta)Y_{F}+\eta\,k_{j}(\eta)$, and
	\eqref{eq:weakdual} yields
	\begin{equation*}
		\bar{Y}_{\infty}(\lambda_{1})\leq Y_{F}
		+\eta\,\bigl[\lambda_{1}k_{1}(\eta)+(1-\lambda_{1})k_{0}(\eta)-1\bigr].
	\end{equation*}
	The term in brackets converges to $\lambda_{1}/a-1<0$ as $\eta\to0$.
	
	\medskip
	\noindent\textit{(b) $Y_{\sigma}(\lambda_{2})\to Y_{F}$ as $\sigma\to\infty$.}
	Let $x(A)\equiv A^{m}/\mathbb{E}[A^{m}]$. The input choices $K_{0}=0$, $N_{0}(A)=x(A)/b$,
	$K_{1}(A)=x(A)/\lambda_{2}$ and $N_{1}(A)=(1-a/\lambda_{2})\,x(A)/b$ are feasible at
	$\lambda_{2}$ because $\lambda_{2}\geq a$. Their aggregate revenue is
	$\bigl[\lambda_{2}\,G(\frac{1}{\lambda_{2}},\frac{1-a/\lambda_{2}}{b})^{\theta}
	+(1-\lambda_{2})\,G(0,\frac{1}{b})^{\theta}\bigr]Y_{F}$,
	which converges to $Y_{F}$ as $\gamma\to1$. By Lemma \ref{lem:duality},
	$Y_{\sigma}(\lambda_{2})$ is at least this revenue, and
	$Y_{\sigma}(\lambda_{2})\leq Y_{\sigma}(1)=Y_{F}$.
	
	\medskip
	\noindent\textit{(c) Conclusion.}
	Since $G\leq G_{\infty}$, the equilibrium input choices at $\lambda_{1}$ yield weakly more
	revenue under $G_{\infty}$, so $Y_{\sigma}(\lambda_{1})\leq\bar{Y}_{\infty}(\lambda_{1})$.
	Together with \eqref{eq:opt:chord}, this implies
	\begin{equation*}
		V_{\sigma}(\lambda_{1})
		\geq\lambda_{1}\,\frac{Y_{\sigma}(\lambda_{2})-\bar{Y}_{\infty}(\lambda_{1})}{\lambda_{2}-\lambda_{1}}
		\qquad\text{and}\qquad
		V_{\sigma}(1)\leq\frac{Y_{F}-Y_{\sigma}(\lambda_{2})}{1-\lambda_{2}}.
	\end{equation*}
	By (a) and (b), the first bound converges to a strictly positive number and the second
	bound converges to zero as $\sigma\to\infty$. Hence there exists
	$\sigma_{H}>\frac{1}{1-\theta}$ such that $V_{\sigma}(\lambda_{1})>V_{\sigma}(1)$ for all
	$\sigma>\sigma_{H}$, so $\lambda^{*}<1$ by (i).
	
	\bigskip
	\noindent\textbf{Part 4:} Existence of $\sigma_{L}$.
	\\
	\\
	Let $\sigma<1$, $G_{0}(K,N)\equiv\min\{K,N\}$, $m_{0}\equiv(\mathbb{E}[A])^{m}$,
	$m_{1}\equiv\mathbb{E}[A^{m}]$ and
	$\hat{Y}(\lambda)\equiv[m_{0}+\lambda(m_{1}-m_{0})]^{1/m}$. Note that $\hat{Y}(1)=Y_{F}$
	and $\hat{Y}'(1)=(1-\theta)Y_{F}-(1-\theta)m_{0}m_{1}^{-\theta}$.
	
	\medskip
	\noindent\textit{(a) $Y_{\sigma}(\lambda)\leq\kappa_{\sigma}^{\theta}\,\hat{Y}(\lambda)$.}
	Under $G_{0}$ and at the prices $w=0$ and $r>0$, firms set $N=K$, so
	$\tilde{\pi}_{1}(0,r)=(1-\theta)\theta^{t}r^{-t}m_{1}$ and
	$\tilde{\pi}_{0}(0,r)=(1-\theta)\theta^{t}r^{-t}m_{0}$. For $r=\theta\hat{Y}(\lambda)$,
	the bound \eqref{eq:weakdual} equals $\hat{Y}(\lambda)$. Any input choices that are
	feasible at $\lambda$ thus yield at most $\hat{Y}(\lambda)$ under $G_{0}$, and by
	\eqref{eq:sandwich} at most $\kappa_{\sigma}^{\theta}\,\hat{Y}(\lambda)$ under $G$.
	
	\medskip
	\noindent\textit{(b) A bound on $V_{\sigma}(1)$.}
	Let $\psi_{0}$ denote the profit of an uninformed firm under $G_{0}$ at the prices
	$(w(1),r(1))$, which do not depend on $\sigma$. Since $G\geq G_{0}$ by
	\eqref{eq:sandwich}, $\pi_{0}(1)\geq\psi_{0}$, and thus
	$V_{\sigma}(1)\leq(1-\theta)Y_{F}-\psi_{0}$ by \eqref{eq:opt:V}. Moreover,
	$\psi_{0}>(1-\theta)m_{0}m_{1}^{-\theta}$: The right-hand side is the maximum of
	$\mathbb{E}[A]K^{\theta}-\theta Y_{F}K$, which is attained at some $\hat{K}>0$. Since
	$w(1)+r(1)=\theta Y_{F}$, an uninformed firm earns exactly this amount under $G_{0}$ if
	it chooses $\hat{K}$ and hires $N=\hat{K}$ in every state. It earns strictly more if it
	hires $N(A)=\min\{\hat{K},(\theta A/w(1))^{m}\}$, because $N(A)<\hat{K}$ with strictly
	positive probability given that the support of $F$ contains zero.
	
	\medskip
	\noindent\textit{(c) Conclusion.}
	By \eqref{eq:opt:chord} and (a), for every $\lambda_{1}<1$,
	\begin{equation*}
		V_{\sigma}(\lambda_{1})
		\geq\lambda_{1}\,\frac{Y_{F}-\kappa_{\sigma}^{\theta}\,\hat{Y}(\lambda_{1})}{1-\lambda_{1}}.
	\end{equation*}
	As $\sigma\to0$, the right-hand side converges to
	$\lambda_{1}\frac{\hat{Y}(1)-\hat{Y}(\lambda_{1})}{1-\lambda_{1}}$ by (P3), and this
	expression converges to $\hat{Y}'(1)$ as $\lambda_{1}\to1$. By (b),
	$\hat{Y}'(1)>(1-\theta)Y_{F}-\psi_{0}\geq V_{\sigma}(1)$. Hence there exist
	$\lambda_{1}<1$ and $\sigma_{L}\in(0,1)$ such that
	$V_{\sigma}(\lambda_{1})>V_{\sigma}(1)$ for all $\sigma<\sigma_{L}$, so
	$\lambda^{*}<1$ by (i).
\end{App-Proof}

\noindent
\begin{App-Proof}{Corollary \ref{prop:seller:incomes}}
	Under perfect competition, the price of data is zero. Since $D(1)>0$, all firms then
	purchase data, so the factor prices are $w(1)$ and $r(1)$. If
	$\sigma\in[1,\frac{1}{1-\theta}]$, then $\lambda^{*}=1$ by Proposition
	\ref{prop:endog:optimal}, so the factor prices are the same in both cases. Now let
	$\sigma<\sigma_{L}$ or $\sigma>\sigma_{H}$, so that $\lambda^{*}<1$. Since $\lambda^{*}$
	maximizes $V$ and $Y$ is strictly increasing, \eqref{eq:opt:V} implies
	$\pi_{0}(\lambda^{*})-\pi_{0}(1)\leq(1-\theta)[Y(\lambda^{*})-Y(1)]<0$. Since
	$\pi_{0}(w,r)$ is decreasing in both prices, $w(\lambda^{*})>w(1)$ or
	$r(\lambda^{*})>r(1)$. If $\sigma<\sigma_{L}$, then $w(\lambda^{*})<w(1)$ because
	$\sigma<1$ (Proposition \ref{prop:exog:factorincome}), and hence
	$r(\lambda^{*})>r(1)$. If $\sigma>\sigma_{H}$, then $r(\lambda^{*})<r(1)$ because
	$\sigma>1$ (Proposition \ref{prop:exog:factorincome}), and hence
	$w(\lambda^{*})>w(1)$.
\end{App-Proof}

\newpage
	\bibliographystyle{apacite}
	
	\bibliography{DI_references_main}
	
	\newpage
	\begin{center}
		{\Large {\bf Data and Inequality} \\ \bigskip {\it Online Appendix}}  \\
		\bigskip
		{\large\begin{tabular}{c c c}
				Carl-Christian Groh
		\end{tabular}}
	\end{center}

	\bigskip \bigskip
	

	\startcontents[sections]
	\printcontents[sections]{l}{1}{\setcounter{tocdepth}{2}}

	\setcounter{section}{1}
	\setcounter{page}{1}
	\setcounter{equation}{0}
	\setcounter{figure}{0}
	\numberwithin{equation}{section}
	\numberwithin{figure}{section}
	
	\section{Omitted Proofs}
	\label{sec:supmat:equilibrium}

	\noindent
	\begin{App-Proof}{Lemma \ref{lem:equilibrium}}
		\textit{(P1).} Every firm hires labor knowing its profitability. Since revenue is
		strictly concave in $(K,N)$, optimal input choices are unique and continuous in prices,
		and they satisfy the stated first-order conditions. By \eqref{eq:focN}, $N_{0}(A)$ is
		strictly increasing in $A$.
		
		\textit{(P2).} Revenue is homogeneous of degree $\theta$ in $(K,N)$. By (P1), the
		expected factor payments of each firm thus equal $\theta$ times its expected revenue,
		and hence $t$ times its expected profit. Moreover, marginal revenues are homogeneous of
		degree $\theta-1$ in $(K,N)$. If both prices are multiplied by $s>0$ and all inputs by
		$s^{-m}$, the first-order conditions in (P1) thus continue to hold, and profits change
		by the factor $s^{-t}$.
		
		\textit{(P3).} Since $\gamma<0$,
		$\min\{a,b\}\min\{K,N\}^{\gamma}\leq aK^{\gamma}+bN^{\gamma}\leq\min\{K,N\}^{\gamma}$.
		Raising all three terms to the power $1/\gamma<0$ yields \eqref{eq:sandwich}.
		
		\textit{Equilibrium conditions.} The remaining steps use Lemma \ref{lem:intensity}, whose
		proof relies only on (P1). For $\rho>0$, let $K_{j}(\rho)$ and $N_{j}(\rho)$ denote the
		expected capital and labor of a firm of type $j\in\{0,1\}$ at the prices
		$(w,r)=(1,\rho)$. By (P2), a firm of type $j$ demands $w^{-m}K_{j}(\rho)$ and
		$w^{-m}N_{j}(\rho)$ at the prices $(w,w\rho)$. Hence these are equilibrium prices for
		$\lambda$ if and only if $w^{m}=\lambda N_{1}(\rho)+(1-\lambda)N_{0}(\rho)$ and
		\begin{equation}\label{eq:excess}
			e(\rho,\lambda)\equiv\lambda\,[K_{1}(\rho)-N_{1}(\rho)]+(1-\lambda)\,[K_{0}(\rho)-N_{0}(\rho)]=0.
		\end{equation}
		
		\textit{Existence.} If $\lambda=1$, \eqref{eq:intensity} implies that
		\eqref{eq:excess} holds if and only if $\rho=\rho_{1}\equiv a/b$. If $\lambda=0$, every
		firm holds $K_{0}=1$. By \eqref{eq:focN}, the expected labor of a firm with $K_{0}=1$ is
		continuous and strictly decreasing in $w$ with limits $\infty$ and $0$, so exactly one
		wage clears the labor market, and \eqref{eq:focK0} yields the unique rental rate at
		which $K_{0}=1$ is optimal. Let $\rho_{0}$ denote the resulting relative price. If
		$\lambda\in(0,1)$ and $\sigma>1$, Lemma \ref{lem:intensity} implies $N_{1}(\rho_{0})<K_{1}(\rho_{0})$ and
		$N_{0}(\rho_{1})>K_{0}(\rho_{1})$, so $e(\rho_{0},\lambda)>0>e(\rho_{1},\lambda)$. If
		$\sigma<1$, both inequalities are reversed, and if $\sigma=1$, then
		$e(\rho_{1},\lambda)=0$. Since $e$ is continuous, \eqref{eq:excess} has a solution
		between $\rho_{0}$ and $\rho_{1}$.
		
		\textit{Uniqueness.} Let $\lambda>0$. By Lemma \ref{lem:duality}, equilibrium prices
		attain the minimum in \eqref{eq:dual}. The profit function $\pi_{0}$ is convex, and
		$\pi_{1}(w,r)=(1-\theta)\theta^{t}\,\mathbb{E}[A^{m}]\,c(w,r)^{-t}$, where $c$ is the
		unit cost function of $G$. Since $c$ is concave, homogeneous of degree one, and
		strictly concave along every line that does not pass through the origin, $c^{-t}$ is
		strictly convex. The objective in \eqref{eq:dual} is thus strictly convex and has a
		unique minimizer. For $\lambda=0$, uniqueness was shown above.
		
		\textit{Continuity.} By the previous steps, the equilibrium value $\rho(\lambda)$ is
		the unique solution to \eqref{eq:excess}, and it lies between $\rho_{0}$ and
		$\rho_{1}$. Since $e$ is continuous in $(\rho,\lambda)$, every limit point of
		$\rho(\lambda_{n})$ for a sequence $\lambda_{n}\to\lambda$ solves \eqref{eq:excess} for
		$\lambda$ and thus equals $\rho(\lambda)$. Hence $\rho(\lambda)$ is continuous, and so
		are $w(\lambda)=[\lambda N_{1}(\rho(\lambda))+(1-\lambda)N_{0}(\rho(\lambda))]^{1/m}$,
		$r(\lambda)=\rho(\lambda)w(\lambda)$, the profits, and $Y(\lambda)$.
	\end{App-Proof}

	\section{Extensions}
	
	\subsection{Noisy Signals}
	\label{sec:supmat:noisy}
	
	\textbf{Model:}
	I now consider an extension that modifies the baseline model from Section \ref{sec:model} in two ways. First, profitability
	is log-normally distributed, i.e., $\ln A\sim\mathcal{N}(\mu,v)$. Second, every firm
	observes a signal $s=\ln A+\varepsilon$ before it chooses capital, where the noise
	$\varepsilon\sim\mathcal{N}(0,v_{\varepsilon})$ is independent of $A$ and across firms. As
	before, all firms know their profitability when they hire labor. The precision of data is
	$\kappa\equiv v/(v+v_{\varepsilon})\in[0,1]$, i.e., the share of the variance of $\ln A$
	that the signal explains. Thus, $\ln A=x+u$, where the forecast
	$x\equiv\mathbb{E}[\ln A\mid s]\sim\mathcal{N}(\mu,\kappa v)$ and the forecast error
	$u\sim\mathcal{N}(0,(1-\kappa)v)$ are independent. If $\kappa=1$, all firms are informed
	as in the baseline model, and if $\kappa=0$, all firms are uninformed.
	
	Given the factor prices $(w,r)$, a firm chooses capital knowing its signal and labor
	knowing its profitability. Its expected profit is
	\begin{equation}\label{eq:ext:profit}
		\pi_{\kappa}(w,r)=\mathbb{E}\biggl[\,\max_{K\geq 0}\;\Bigl\{
		\mathbb{E}\Bigl[\,\max_{N\geq 0}\bigl\{A\,G(K,N)^{\theta}-wN\bigr\}\Bigm|s\Bigr]-rK\Bigr\}\biggr].
	\end{equation}
	An equilibrium consists of prices $(w,r)$ and input choices such that all firms choose
	their inputs optimally and both factor markets clear, i.e., expected capital and expected
	labor per firm both equal one. I denote the equilibrium wage, rental rate and total income
	by $w(\kappa)$, $r(\kappa)$ and $Y(\kappa)$. I refer to this model as the noisy signals model.
	\\
	\\
	\textbf{Analysis:}
	I establish that equivalent versions of the results in Propositions \ref{prop:exog:factorincome} and \ref{prop:exog:profits} emerge in the noisy signals model. 
	
	\begin{proposition}[Noisy Signals]
		\label{prop:ext:imprecise}
		\hfill \break
		Consider the noisy signals model.
		\begin{itemize}
			\item The equilibrium wage is strictly increasing in $\kappa$ if
			$\sigma<\frac{1}{1-\theta}$ and strictly decreasing in $\kappa$ if
			$\sigma>\frac{1}{1-\theta}$.
			\item If $\sigma\geq1$, the equilibrium rental rate is strictly increasing in $\kappa$.
			\item Total profits are strictly increasing in $\kappa$. The gap between the expected
			profits that a firm with perfect data and a firm without data would earn at the
			equilibrium factor prices is positive and strictly decreasing in $\kappa$.
		\end{itemize}
	\end{proposition}
	\noindent
	More precise data thus affect the wage, the rental rate and profits in the same way as a
	larger share of informed firms in the baseline model. The result for the wage is sharper:
	the wage is monotone in the precision of data for every elasticity of substitution, so
	$\frac{1}{1-\theta}$ is the exact threshold at which the effect of data on the wage
	changes sign. The reason is that an economy with imprecise data behaves like an economy
	without data in which the variance of profitability is $(1-\kappa)v$. More precise data
	reduce the uncertainty that remains when firms choose capital, and only this residual
	uncertainty matters.
	
	\subsection{Elastic Factor Demands and Costly Data}
	
	\label{sec:supmat:elastic}
	
	\noindent
	\textbf{Model:}
	I now consider an extension that modifies the baseline model in two ways: First, the supplies of labor and
	capital are no longer fixed. At the wage $w$, workers supply $N^{s}(w)=w^{\eta_{N}}$ units of
	labor, and at the rental rate $r$, capital owners supply $K^{s}(r)=r^{\eta_{K}}$ units of
	capital, where $\eta_{N}\geq0$ and $\eta_{K}\geq0$ are the elasticities of labor supply
	and capital supply. Second, the data provider incurs a cost $c\geq0$ for every firm that
	purchases data. The baseline model is the special case $\eta_{N}=\eta_{K}=c=0$.
	
	The share of informed firms is exogenously given by $\lambda$.  An equilibrium consists of prices $(w,r)$
	and input choices such that all firms choose their inputs optimally and both factor
	markets clear, i.e., the labor and capital that firms demand equal $N^{s}(w)$ and $K^{s}(r)$. As
	before, I denote the equilibrium prices by $w(\lambda)$ and $r(\lambda)$, expected profits
	by $\pi_{1}(\lambda)$ and $\pi_{0}(\lambda)$, total income by $Y(\lambda)$, and the value of information by $D(\lambda)\equiv\pi_{1}(\lambda)-\pi_{0}(\lambda)$. I assume that $c<D(0)$, which
	ensures that a positive share of firms purchases data in equilibrium. The surplus of workers and
	capital owners is the area above their supply curves, so total surplus gross of the costs
	of data is
	\begin{equation}\label{eq:elastic:welfare}
		W(\lambda)\;\equiv\;(1-\theta)\,Y(\lambda)
		+\frac{w(\lambda)^{1+\eta_{N}}}{1+\eta_{N}}
		+\frac{r(\lambda)^{1+\eta_{K}}}{1+\eta_{K}}.
	\end{equation}
	\noindent
	\textbf{Analysis:}
	We first establish that the results regarding the effect of data on the wage rate and the value of information extend:
	I first establish that the results regarding the effect of data on the wage and the value
	of data extend:
	
	\begin{proposition}[Elastic Factor Supplies]
		\label{prop:elastic:wage}
		\hfill \break
		Consider the model with elastic factor supplies, and let $\eta_{N},\eta_{K}\geq0$.
		\begin{itemize}
			\item If $\sigma\leq1$, the equilibrium wage is strictly increasing in $\lambda$. If
			$\sigma>1$, then $w(1)>w(0)$ if and only if $\sigma<\frac{1}{1-\theta}$.
			\item The profit gap $D(\lambda)$ is positive and strictly decreasing in $\lambda$.
			\item The rental rate is strictly increasing in $\lambda$ if $\sigma > 1$.
		\end{itemize}
		If further $\eta_K = \eta_N$, total profits are increasing in $\lambda$. 
	\end{proposition}
	\noindent
	As in the baseline analysis, increases in the number of informed firms raise the equilibrium wage if capital and labor are sufficiently complementary, and decrease the value of information. Further, improvements in the availability of data raise the rental rate if $\sigma$ is large. The effects of increases in data availability on the rental rate at low $\sigma$ depend on the spread of the profitability distribution, as in the baseline analysis. The intuition underlying these results is as in the baseline analysis. The results on total profits, which were somewhat secondary in the baseline analysis, carry over if both elasticities are the same.  If labor supply is much more elastic than capital supply and
	$\sigma>\frac{1}{1-\theta}$, informed firms replace labor with capital, so employment and
	total income can fall in $\lambda$ even though total surplus rises, with the gains accruing
	to capital owners through a higher rental rate.

	\section{Proofs: Extensions}
	
	\noindent
	\begin{App-Proof}{Proposition \ref{prop:ext:imprecise}}
		Let $\tau\equiv(1-\kappa)v$, $X\equiv e^{x}$ and $U_{\tau}\equiv e^{\sqrt{\tau}\,z}$,
		where $z$ is a standard normal random variable that is independent of $x$, so that
		$A=XU_{\tau}$. Let $\tilde g(y)\equiv G(y,1)$,
		$\psi(y)\equiv\tilde g(y)^{(\gamma-\theta)m}$ and $\varphi(n)\equiv n\,\psi(1/n)$.
		
		\bigskip
		\noindent\textbf{Part 1: Preliminaries.}
		
		\medskip
		\noindent\textit{(i) An equivalent economy without data.}
		Fix prices $w,r>0$. Since revenue is homogeneous of degree $\theta$ in $(K,N)$, the
		objective of a firm with forecast $x$ at the inputs $(X^{m}K,X^{m}N)$ equals $X^{m}$
		times the objective of an uninformed firm with profitability $U_{\tau}$ at the inputs
		$(K,N)$. A firm with forecast $x$ thus chooses $X^{m}$ times the inputs
		$(k,n(U_{\tau}))$ of this uninformed firm, and both factor markets clear if and only if
		$\mathbb{E}[X^{m}]\,k=\mathbb{E}[X^{m}]\,\mathbb{E}[n(U_{\tau})]=1$. Let
		$B\equiv(\mathbb{E}[X^{m}])^{1/m}=e^{\mu+m\kappa v/2}$. An uninformed firm with
		profitability $BU_{\tau}$ chooses the inputs $B^{m}(k,n(U_{\tau}))$ and earns the
		expected profit $\pi_{\kappa}(w,r)$. Hence $(w,r)$ are equilibrium prices if and only if
		they are equilibrium prices of the economy without data in which profitability is
		$BU_{\tau}$. By Lemma \ref{lem:equilibrium}, the equilibrium exists and is unique. In an
		economy without data, the allocation of labor does not depend on the scale of
		profitability, and the factor prices are proportional to it. Therefore,
		\begin{equation}\label{eq:ext:scaling}
			w(\kappa)=e^{\mu+m\kappa v/2}\,\omega(\tau)
			\qquad\text{and}\qquad
			r(\kappa)=e^{\mu+m\kappa v/2}\,\omega(\tau)\,\rho(\kappa),
		\end{equation}
		where $\omega(\tau)$ is the wage in the economy without data in which profitability is
		$U_{\tau}$. Let $n_{\tau}$ denote the labor of a firm in this economy, so that
		$\mathbb{E}[n_{\tau}]=1$ and $n_{0}=1$.
		
		\medskip
		\noindent\textit{(ii) Dispersion.}
		Let $h$ be a strictly increasing function and let $Z_{\tau}\equiv h(\sqrt{\tau}\,z-c_{\tau})$
		for $\tau\geq0$, where the constants $c_{\tau}$ are such that $\mathbb{E}[Z_{\tau}]$ does
		not depend on $\tau$. I show that $\mathbb{E}[f(Z_{\tau})]$ is strictly increasing in
		$\tau$ for every strictly convex function $f$. Let $\tau<\tau'$. Since
		$(\sqrt{\tau'}-\sqrt{\tau})z-(c_{\tau'}-c_{\tau})$ is strictly increasing in $z$ and
		$\mathbb{E}[Z_{\tau'}]=\mathbb{E}[Z_{\tau}]$, there exists $z_{0}$ such that
		$Z_{\tau'}>Z_{\tau}$ if $z>z_{0}$ and $Z_{\tau'}<Z_{\tau}$ if $z<z_{0}$. Let $d$ be the
		slope of a supporting line of $f$ at $h(\sqrt{\tau}\,z_{0}-c_{\tau})$. Since $Z_{\tau}$ is
		weakly increasing in $z$ and $f$ is strictly convex,
		$f(Z_{\tau'})-f(Z_{\tau})>d\,(Z_{\tau'}-Z_{\tau})$ for all $z\neq z_{0}$. Taking
		expectations yields $\mathbb{E}[f(Z_{\tau'})]>\mathbb{E}[f(Z_{\tau})]$.
		
		\medskip
		\noindent\textit{(iii) Labor.}
		Let $\nu(\ell)$ denote the solution $n$ to
		$\theta b\,e^{\ell}n^{\gamma-1}G(1,n)^{\theta-\gamma}=1$, which is strictly increasing
		in $\ell$. By \eqref{eq:focN}, a firm with capital $K$ and profitability $U_{\tau}$
		hires $K\,\nu(\sqrt{\tau}\,z-c)$ units of labor, where $c\equiv\ln(wK^{1-\theta})$. In
		particular, $n_{\tau}=\nu(\sqrt{\tau}\,z-\ln\omega(\tau))$.
		
		\bigskip
		\noindent\textbf{Part 2: The wage.}
		By \eqref{eq:focN} and $G(K,N)=N\,\tilde g(K/N)$, every firm satisfies
		$A^{m}=(w/(\theta b))^{m}\,N\,\psi(K/N)$ in every state. In the economy without data in
		which profitability is $U_{\tau}$, every firm holds one unit of capital, so taking
		expectations yields
		$e^{m^{2}\tau/2}=\mathbb{E}[U_{\tau}^{m}]=(\omega(\tau)/(\theta b))^{m}\,\mathbb{E}[\varphi(n_{\tau})]$.
		Since $w(1)=\theta b\,e^{\mu+mv/2}$ by \eqref{eq:fullinfo}, \eqref{eq:ext:scaling} implies
		\begin{equation}\label{eq:ext:wage}
			\Bigl(\frac{w(\kappa)}{w(1)}\Bigr)^{m}=\frac{1}{\mathbb{E}[\varphi(n_{\tau})]}.
		\end{equation}
		The function $\tilde g$ is strictly increasing and strictly concave, and
		$\varphi''(n)=\psi''(1/n)/n^{3}$. If $\gamma<\theta$, then $\psi$ is a strictly
		decreasing and strictly convex transformation of $\tilde g$, so $\varphi$ is strictly
		convex. If $\gamma>\theta$, then $\psi$ is a strictly increasing and strictly concave
		transformation of $\tilde g$, so $\varphi$ is strictly concave. By (ii) and (iii),
		$\mathbb{E}[\varphi(n_{\tau})]$ is thus strictly increasing in $\tau$ if
		$\sigma<\frac{1}{1-\theta}$ and strictly decreasing in $\tau$ if
		$\sigma>\frac{1}{1-\theta}$. Since $\tau$ is strictly decreasing in $\kappa$,
		\eqref{eq:ext:wage} yields the first claim.
		
		\bigskip
		\noindent\textbf{Part 3: The rental rate and total profits.}
		By \eqref{eq:intensity}, $\rho(\kappa)=\frac{a}{b}\,\mathbb{E}[n_{\tau}^{1/\sigma}]$. By
		(ii) and (iii), $\rho(\kappa)$ is strictly increasing in $\kappa$ if $\sigma>1$,
		constant if $\sigma=1$, and strictly decreasing if $\sigma<1$. Moreover, $\rho(\kappa)$
		is continuous, because $\omega(\tau)$ is the unique solution to
		$\mathbb{E}[\nu(\sqrt{\tau}\,z-\ln\omega)]=1$ and this expectation is continuous in
		$(\tau,\omega)$.
		
		By (i) and Lemma \ref{lem:duality},
		$Y(\kappa)=\min_{w,r>0}\{\pi_{\kappa}(w,r)+w+r\}$, and the equilibrium prices attain the
		minimum. Let $\kappa<\kappa'$. A firm with precision $\kappa'$ can add noise to its
		signal and then choose the capital of a firm with precision $\kappa$. This choice is
		not optimal, because revenue is strictly concave and the optimal capital of the firm
		varies with its forecast. Hence $\pi_{\kappa}(w,r)<\pi_{\kappa'}(w,r)$ at any prices, and
		evaluating $\pi_{\kappa}(w,r)+w+r$ at the equilibrium prices for $\kappa'$ yields
		$Y(\kappa)<Y(\kappa')$. By (P2), total profits equal $(1-\theta)Y(\kappa)$ and
		$w(\kappa)+r(\kappa)=\theta\,Y(\kappa)$. Total profits are thus strictly increasing in
		$\kappa$, and so is $r(\kappa)=\theta\,Y(\kappa)\rho(\kappa)/(1+\rho(\kappa))$ if
		$\sigma\geq1$.
		
		\bigskip
		\noindent\textbf{Part 4: The profit gap.}
		Let $\pi_{1}(w,r)$ and $\pi_{0}(w,r)$ denote the expected profits of a firm with perfect
		data and of a firm without data. As in the proof of Proposition
		\ref{prop:exog:profits}, $\pi_{1}>\pi_{0}$ at any prices. By (P2),
		$\pi_{j}(w,r)=w^{-t}g_{j}(\rho)$ with $g_{j}(\rho)\equiv\pi_{j}(1,\rho)$, and
		$w(\kappa)=\theta\,Y(\kappa)/(1+\rho(\kappa))$. The profit gap at the equilibrium prices
		is therefore
		\begin{equation}\label{eq:ext:gap}
			\bigl(\theta\,Y(\kappa)\bigr)^{-t}\,H(\rho(\kappa)),
			\qquad\text{where}\qquad
			H(\rho)\equiv(1+\rho)^{t}\,[g_{1}(\rho)-g_{0}(\rho)].
		\end{equation}
		Let $s_{1}$ and $s_{0}$ denote the shares of capital in the expected factor payments of
		these two firms, and let $\bar s\equiv\rho/(1+\rho)$. By Hotelling's lemma and (P2),
		$\rho\,g_{j}'(\rho)=-t\,s_{j}\,g_{j}(\rho)$, so
		\begin{equation*}
			H'(\rho)=\frac{t}{\rho}\,(1+\rho)^{t}\,\bigl[(\bar s-s_{1})\,g_{1}(\rho)-(\bar s-s_{0})\,g_{0}(\rho)\bigr].
		\end{equation*}
		I next compare firms whose forecast errors have different variances $\tau_{i}$ at given
		prices. By \eqref{eq:intensity}, the ratio $q_{i}$ of labor to capital of such a firm
		satisfies $\mathbb{E}[q_{i}^{1/\sigma}]=\frac{b}{a}\rho$ for every $\tau_{i}$, and by
		(iii), $q_{i}=\nu(\sqrt{\tau_{i}}\,z-c_{i})$. Applying (ii) to $q_{i}^{1/\sigma}$ and
		the function $f(y)=y^{\sigma}$ shows that $\mathbb{E}[q_{i}]$ is strictly increasing in
		$\tau_{i}$ if $\sigma>1$ and strictly decreasing if $\sigma<1$. The share of capital in
		the factor payments of a firm is thus strictly decreasing in $\tau_{i}$ if $\sigma>1$
		and strictly increasing if $\sigma<1$. At the equilibrium prices for $\kappa$, this
		share equals $\bar s$ for the firms in the economy by market clearing, and these
		firms have $\tau_{i}=\tau\in[0,v]$, whereas $\tau_{i}=0$ for a firm with perfect data
		and $\tau_{i}=v$ for a firm without data. Hence, at $\rho=\rho(\kappa)$,
		$s_{1}\geq\bar s\geq s_{0}$ with at least one strict inequality if $\sigma>1$, and the
		reverse if $\sigma<1$. Consequently, $H'(\rho(\kappa))<0$ for all $\kappa$ if
		$\sigma>1$, and $H'(\rho(\kappa))>0$ for all $\kappa$ if $\sigma<1$. Since
		$\rho(\kappa)$ is continuous and strictly monotone as shown in Part 3,
		$H(\rho(\kappa))$ is strictly decreasing in $\kappa$ in both cases, and it is constant
		if $\sigma=1$. Since $Y(\kappa)$ is strictly increasing, \eqref{eq:ext:gap} is strictly
		decreasing in $\kappa$.
	\end{App-Proof}
	
	
	\noindent
	\begin{App-Proof}{Proposition \ref{prop:elastic:wage}}
		For $\lambda\in[0,1]$ and prices $w,r>0$, let
		$N^{d}_{\lambda}(w,r)\equiv\lambda N_{1}+(1-\lambda)N_{0}$ and
		$K^{d}_{\lambda}(w,r)\equiv\lambda K_{1}+(1-\lambda)K_{0}$ denote the aggregate factor
		demands, and let
		\begin{equation}\label{eq:el:Phi}
			\Phi_{\lambda}(w,r)\equiv\lambda\,\pi_{1}(w,r)+(1-\lambda)\,\pi_{0}(w,r)
			+\frac{w^{1+\eta_{N}}}{1+\eta_{N}}+\frac{r^{1+\eta_{K}}}{1+\eta_{K}}.
		\end{equation}
		Throughout, $\lambda<\lambda'$, and $(w,r)$ and $(w',r')$ denote the equilibrium prices
		for $\lambda$ and $\lambda'$.
		
		\bigskip
		\noindent\textbf{Part 1: Preliminaries.}
		
		\medskip
		\noindent\textit{(i) Equilibrium.}
		Optimal input choices are unique, so $\pi_{1}$ and $\pi_{0}$ are differentiable, and
		by Hotelling's lemma,
		\begin{equation}\label{eq:el:grad}
			\nabla\Phi_{\lambda}(w,r)=\bigl(N^{s}(w)-N^{d}_{\lambda}(w,r),\;K^{s}(r)-K^{d}_{\lambda}(w,r)\bigr).
		\end{equation}
		Moreover, $\pi_{1}$ and $\pi_{0}$ are strictly convex, because by (P1), input choices
		that are optimal at some prices are not optimal at any other prices. Hence
		$\Phi_{\lambda}$ is strictly convex, and prices are equilibrium prices for $\lambda$ if
		and only if they minimize $\Phi_{\lambda}$. A minimizer exists, because
		$\Phi_{\lambda}$ is continuous, tends to infinity as $w$ or $r$ tends to infinity, and
		is decreasing in each price near zero, since the demand for each factor is unbounded
		as its price vanishes. The equilibrium thus exists and is unique.
		
		\medskip
		\noindent\textit{(ii) Fixed supplies.}
		Let $\bar N\equiv N^{s}(w)$ and $\bar K\equiv K^{s}(r)$, and consider the economy in
		which the supplies of labor and capital are fixed at these values. After a change of
		units, this economy is a baseline model with a different weight $a$ and a rescaled
		profitability, so Lemmas \ref{lem:duality} and \ref{lem:equilibrium} and Proposition
		\ref{prop:exog:factorincome} apply to it. Its equilibrium prices for $\lambda$ are
		$(w,r)$. Let $(w_{b},r_{b})$ denote its equilibrium prices for $\lambda'$. They minimize
		the strictly convex function
		$\Phi^{b}(\hat w,\hat r)\equiv\lambda'\pi_{1}(\hat w,\hat r)+(1-\lambda')\pi_{0}(\hat w,\hat r)+\hat w\bar N+\hat r\bar K$,
		whose gradient is
		$(\bar N-N^{d}_{\lambda'}(\hat w,\hat r),\,\bar K-K^{d}_{\lambda'}(\hat w,\hat r))$.
		
		\medskip
		\noindent\textit{(iii) Factor demands.}
		At any prices, $\pi_{1}>\pi_{0}$ as in the proof of Proposition
		\ref{prop:exog:profits}, and by (P2), $wN_{j}=(1-s_{j})\,t\,\pi_{j}$ and
		$rK_{j}=s_{j}\,t\,\pi_{j}$ for $j\in\{0,1\}$. By (P3), $N_{1}>N_{0}$ if $\sigma\leq1$
		and $K_{1}>K_{0}$ if $\sigma\geq1$. Next, fix the wage. The left-hand side of
		\eqref{eq:focN} is strictly decreasing in $N$, and it is increasing in $K$ if
		$\sigma<m$ and decreasing in $K$ if $\sigma>m$. The labor of a firm is thus increasing
		in its capital if $\sigma<m$ and decreasing in its capital if $\sigma>m$. Since the
		capital of every firm is decreasing in $r$, both $N_{1}$ and $N_{0}$ are weakly
		decreasing in $r$ if $\sigma\leq m$, and both are weakly increasing in $r$ if
		$\sigma\geq m$. Analogously, at a given rental rate, both $K_{1}$ and $K_{0}$ are
		weakly decreasing in $w$ if $\sigma\leq m$ and weakly increasing in $w$ if
		$\sigma\geq m$, because a higher wage lowers the labor of every firm for given
		capital, which lowers the marginal revenue of capital if $\sigma<m$ and raises it if
		$\sigma>m$.
		
		\medskip
		\noindent\textit{(iv) Comparison.}
		I show that $w'>w$ if $w_{b}>w$ and $N_{1}>N_{0}$ at the prices $(w,r)$. Since
		$N^{d}_{\lambda}(w,r)=\bar N$,
		\begin{equation*}
			N^{d}_{\lambda'}(w,r)=\bar N+(\lambda'-\lambda)\,(N_{1}-N_{0})>\bar N.
		\end{equation*}
		Aggregate capital demand $K^{d}_{\lambda'}(w,\cdot)$ is continuous and strictly
		decreasing in the rental rate, with limits $\infty$ and $0$. Let $r_{1}$ and $r_{2}$
		denote the rental rates at which it equals $\bar K$ and $K^{s}(r_{2})$, respectively. If
		$K^{d}_{\lambda'}(w,r)\geq\bar K=K^{s}(r)$, then $r\leq r_{1}$ and
		$K^{d}_{\lambda'}(w,r_{1})=\bar K\leq K^{s}(r_{1})$, so $r\leq r_{2}\leq r_{1}$.
		Otherwise, $r_{1}\leq r_{2}\leq r$ by the same argument. By (ii), the gradient of
		$\Phi^{b}$ at $(w,r_{1})$ is $(\bar N-N^{d}_{\lambda'}(w,r_{1}),0)$. Since $\Phi^{b}$
		is strictly convex and minimized at $(w_{b},r_{b})$,
		\begin{equation*}
			0\;\geq\;\Phi^{b}(w_{b},r_{b})-\Phi^{b}(w,r_{1})\;>\;\bigl[\bar N-N^{d}_{\lambda'}(w,r_{1})\bigr]\,(w_{b}-w),
		\end{equation*}
		so $N^{d}_{\lambda'}(w,r_{1})>\bar N$ because $w_{b}>w$. By (iii),
		$N^{d}_{\lambda'}(w,\cdot)$ is monotone, so $N^{d}_{\lambda'}(w,r_{2})>\bar N$ as well.
		By \eqref{eq:el:grad}, the gradient of $\Phi_{\lambda'}$ at $(w,r_{2})$ is thus
		$(\bar N-N^{d}_{\lambda'}(w,r_{2}),0)$, and its first entry is strictly negative. Since
		$\Phi_{\lambda'}$ is strictly convex and minimized at $(w',r')$,
		\begin{equation*}
			0\;\geq\;\Phi_{\lambda'}(w',r')-\Phi_{\lambda'}(w,r_{2})\;>\;\bigl[\bar N-N^{d}_{\lambda'}(w,r_{2})\bigr]\,(w'-w),
		\end{equation*}
		so $w'>w$. Exchanging the roles of the two factors shows that $r'>r$ if $r_{b}>r$ and
		$K_{1}>K_{0}$ at the prices $(w,r)$.
		
		\bigskip
		\noindent\textbf{Part 2: The profit gap.}
		By (iii), $D(\lambda)>0$. By (i), $\Phi_{\lambda}(w,r)\leq\Phi_{\lambda}(w',r')$ and
		$\Phi_{\lambda'}(w',r')\leq\Phi_{\lambda'}(w,r)$. Adding these inequalities yields
		$(\lambda'-\lambda)\,[D(\lambda')-D(\lambda)]\leq0$. If $D(\lambda')=D(\lambda)$, both
		inequalities bind, so $(w,r)$ minimizes $\Phi_{\lambda'}$ and thus equals $(w',r')$.
		Both factor markets then clear at the same prices for $\lambda$ and $\lambda'$, which
		requires $N_{1}=N_{0}$ and $K_{1}=K_{0}$ and hence $\pi_{1}=\pi_{0}$ by (P2), a
		contradiction. Hence $D(\lambda')<D(\lambda)$.
		
		\bigskip
		\noindent\textbf{Part 3: The wage for $\sigma\leq1$ and the rental rate.}
		By (ii) and Proposition \ref{prop:exog:factorincome}, $w_{b}>w$ if $\sigma\leq1$ and
		$r_{b}>r$ if $\sigma\geq1$. By (iii) and (iv), $w'>w$ if $\sigma\leq1$ and $r'>r$ if
		$\sigma\geq1$.
		
		\bigskip
		\noindent\textbf{Part 4: The wage for $\sigma>1$.}
		For $j\in\{0,1\}$, let $w_{j}$ denote the equilibrium wage for $\lambda=j$, and let
		$\bar N_{j}$ and $\bar K_{j}$ denote the equilibrium supplies of labor and capital. By
		\eqref{eq:fullinfo} and the change of units in (ii), the wage in an economy with fixed
		supplies $(\bar K,\bar N)$ in which all firms are informed is
		$\omega(\bar K,\bar N)\equiv\theta b\,Y_{F}\,\bar N^{\gamma-1}G(\bar K,\bar N)^{\theta-\gamma}$.
		This wage is strictly decreasing in $\bar N$, and it is strictly increasing in $\bar K$ if
		$\sigma<m$ and weakly decreasing in $\bar K$ if $\sigma\geq m$. Moreover,
		$w_{1}=\omega(\bar K_{1},\bar N_{1})$, and by Proposition \ref{prop:exog:factorincome},
		$\omega(\bar K_{0},\bar N_{0})>w_{0}$ if and only if $\sigma<m$. By Part 3,
		$\bar K_{1}\geq\bar K_{0}$. If $\sigma<m$ and $w_{1}\leq w_{0}$, then
		$\bar N_{1}\leq\bar N_{0}$ and thus
		$w_{1}\geq\omega(\bar K_{0},\bar N_{0})>w_{0}$, a contradiction. If $\sigma\geq m$ and
		$w_{1}>w_{0}$, then $\bar N_{1}\geq\bar N_{0}$ and thus
		$w_{1}\leq\omega(\bar K_{0},\bar N_{0})\leq w_{0}$, a contradiction. Hence $w(1)>w(0)$ if
		and only if $\sigma<m$.
		
		\bigskip
		\noindent\textbf{Part 5: Total profits.}
		Suppose that $\eta_{N}=\eta_{K}=\eta$, and let $W(\lambda)\equiv\Phi_{\lambda}(w,r)$.
		By (P2), total profits equal $(1-\theta)Y(\lambda)$ and the factor payments
		$w^{1+\eta}+r^{1+\eta}$ equal $\theta\,Y(\lambda)$, so
		$W(\lambda)=Y(\lambda)\,(1+\eta-\theta\eta)/(1+\eta)$. By (i) and Part 2,
		$W(\lambda)\leq\Phi_{\lambda}(w',r')=W(\lambda')-(\lambda'-\lambda)\,D(\lambda')<W(\lambda')$.
		Hence $Y(\lambda)$ and total profits are strictly increasing in $\lambda$.
	\end{App-Proof}
	
\end{document}